\documentclass{nature_meth}
\usepackage{amssymb,amsfonts,amsmath}
\usepackage{graphicx} 
\usepackage{gensymb}
\usepackage{booktabs}
 \usepackage{multirow}
 \usepackage[flushleft]{threeparttable}
\usepackage{epsfig}
\usepackage[figoff]{figcaps}
\usepackage{hyperref}
\hypersetup{
    colorlinks=true,
    linkcolor=blue,
    filecolor=magenta,
    urlcolor=cyan,
}
\usepackage{adjustbox}
\usepackage[mathlines]{lineno}

\usepackage{outline}
\usepackage{pmgraph}
\usepackage[normalem]{ulem}
\usepackage[utf8]{inputenc}
\usepackage{amssymb}
\usepackage{hyperref}
\usepackage{amsmath}
\usepackage{graphicx}
\usepackage{times}
\usepackage{xcolor}
\usepackage{xspace}
\usepackage[colorinlistoftodos]{todonotes} 
\usepackage{cite}
\usepackage{bm} 
\usepackage{url}

\usepackage{epstopdf}
\AppendGraphicsExtensions{.tiff}
\graphicspath{{fig/}} 

\usepackage{epsfig}
\usepackage{tikz}
\usetikzlibrary{spy}
\usepackage{algpseudocode}
\usepackage{algorithm}
\usepackage{mathrsfs}

\long\def\comment#1{} 

\newcommand{\beq}{\begin{equation}}
\newcommand{\eeq}{\end{equation}}
\newcommand{\beqa}{\begin{eqnarray}}
\newcommand{\eeqa}{\end{eqnarray}}

\usepackage[margin=2.54cm]{geometry}

\usepackage{caption,setspace}
\renewcommand{\spacing}[1]{\renewcommand{\baselinestretch}{#1}}
\spacing{1}

\usepackage{url}

\usepackage{graphicx}
\makeatletter
\let\saved@includegraphics\includegraphics
\AtBeginDocument{\let\includegraphics\saved@includegraphics}
\makeatother

\usepackage{tgadventor}
\usepackage[scaled]{helvet}
 
\usepackage[T1]{fontenc}

\title{Self-supervised Multi-modal Training from Uncurated Image and Reports Enables Zero-shot Oversight Artificial Intelligence in Radiology}

\author{Sangjoon Park$^{\ast1}$,
Eun Sun Lee$^{\dagger,2}$,
Kyung Sook Shin$^{3}$,
Jeong Eun Lee$^{\dagger,3}$, 
and Jong Chul Ye$^{\dagger,\ddagger,4}$
}

\begin{document}
\setstretch{1}

\maketitle
\begin{affiliations}
\item Department of Radiation Oncology, Yonsei College of Medicine, Seoul, Korea
\item Chung-Ang University Hospital, Seoul,  Korea
\item Department of Radiology, Chungnam National University Hospital, Chungnam National University College of Medicine, Daejeon, Korea
\item Kim Jaechul Graduate School of AI, KAIST, Daejeon, Korea
\item[] $^{\ast}$This work was mainly conducted when the first author was affliated with KAIST.
\item[] $^{\dagger}$Co-corresponding authors.
\item[] $^{\ddagger}$Correspondence should be addressed to J.C.Y. (jong.ye@kaist.ac.kr)
\end{affiliations}
 
\begin{abstract}
Oversight AI is an emerging concept in radiology where the AI forms a symbiosis with radiologists by continuously supporting radiologists in their decision-making.
Recent advances in vision-language models  
sheds a light on the long-standing problems of the oversight AI by the understanding  both visual and textual concepts 
and their semantic correspondences. 
However, there have been limited successes in the application of vision-language models in the medical domain, as the current vision-language models and learning strategies for photographic images and captions call for the web-scale data corpus of image and text pairs which was not often feasible in the medical domain. 
To address this, here we present a model dubbed Medical Cross-attention Vision-Language model (Medical X-VL), leveraging the key components to be tailored for the medical domain. Our medical X-VL model is based on the following components: self-supervised uni-modal models in medical domain and fusion encoder to bridge them, momentum distillation, sentence-wise contrastive learning for medical reports, and the sentence similarity-adjusted hard negative mining.
We experimentally demonstrated that our model enables various zero-shot tasks for oversight AI, ranging from the zero-shot classification to zero-shot error correction. Our model outperformed the current state-of-the-art models in two different medical image database, suggesting the novel clinical usage of our oversight AI model for monitoring human errors. Our method was especially successful in the data-limited setting, which is frequently encountered in the clinics, suggesting the potential widespread applicability in medical domain. 
\end{abstract}

\vspace{-0.5cm}
\section*{}
In recent years, deep learning has made significant strides in the development of vision and language models, particularly in the medical field, bringing us closer to human-level intelligence. However, despite these successes, there has been limited progress in building models that can correlate visual and language concepts, unlike the human perception that can seamlessly integrate both modalities. This issue has been a long-standing topic of research in the field of artificial intelligence (AI)\cite{margaret2008mind}. Fortunately, recent advances in vision-language models, a multi-modal model trained on a vast corpus of image-text pairs with the goal of learning shared concepts between images and texts, has led to remarkable results in downstream tasks such as image-text retrieval, vision question answering, visual grounding, and more, which require a deep understanding of both visual and language information. Consequently, the vision-language model has revolutionized the field of multi-modal vision-language research, leading to a significant number of studies in recent years\cite{jia2021scaling, li2021align, cho2021unifying, chen2020uniter, lu2019vilbert, li2020oscar, huang2020pixel}.

The rapid advances of VLP have been indebted to the introduction of vision transformer (ViT)\cite{dosovitskiy2020image}, which processes the images as a set of small patches similar to those of several words for a sentence with the transformer model for natural language processing\cite{vaswani2017attention}. Thanks to the intrinsic similarities between the ways of processing images and sentences through the self-attention mechanism of the transformer, direct attention between the image patches and words are possible, facilitating more straightforward cross-attention between modalities. The recent works have demonstrated that the transformer-based vision-language models trained with the web-scale image and text data pairs have the generic capability for multiple downstream vision-language tasks\cite{radford2021learning, jia2021scaling, yu2022coca, wang2021vlmo, alayrac2022flamingo}. Compared with individual models specialized for each task and modality, the vision-language models trained with massive data exhibit superior performances along with the amortization of training cost, enabling to push the limit of model capacity for both domains to reach human-level performances.

In medical fields, rather than completely replacing clinicians' decision-making tasks, there is also an increasing need for oversight AIs to alert clinically significant abnormalities or to detect and correct rare but critical errors in their clinical decision-making. In particular, the decision of radiologists usually comes in the form of medical reports so that visual language models that can understand both the medical images and the reports are an essential step toward the wide acceptance of an oversight AI.


Thanks to the recent surge of interest in introducing AI into the medical field, there has been a proliferation of self-supervised pre-trained models that are specialized in specific medical domain data and modalities\cite{zhang2022contrastive, huang2021gloria, boecking2022making, tiu2022expert, naseem2022vision}
Furthermore, uncurated medical data such as radiograph image and report pairs are already abundant in hospitals, but the absence of manual annotation to discrete labels for traditional supervised learning impedes the utilization of those uncurated image and text pairs to build a robust model. Therefore, making the model directly learn from the uncurated image-report pairs will greatly increase the usability of data, thereby enabling to develop a robust model that can efficiently adapt to various downstream tasks.
Nevertheless, there exist limited studies on the vision-language models in the medical domain\cite{moon2021multi, yan2022clinical, tiu2022expert, zhang2022contrastive, huang2021gloria, boecking2022making, wu2023medklip} applied for narrow range of tasks, where the pairs of image and sentence are frequently used in the form of radiographs, pathology slides, and corresponding reports, and there is no study proposing the method to bride the uni-modal pre-trained vision and language models in medical domain to build the robust multi-modal model, enabling the zero-shot oversight AI in various applications.


Directly adapting the vision-language model in computer vision to the medical domain may result in suboptimal performance due to the different characteristics between the two domains.
Compared with the photographic images and captions where billion-scale image-text pairs can be utilized with web crawling\cite{radford2021learning, jia2021scaling}, the amount of image-text pairs for medical images is often not sufficient to enable learning a firm relation between visual semantics and textual concept. Furthermore, the diversities between the different images and reports are often subtle in medical domains than photographic images. For radiographs as an example, the standardized imaging protocols make them consistent in anatomical patterns, and the abnormal findings in radiographs are usually subtly different in appearance\cite{xiang2021painting}. 
Likewise, the medical reports usually take the confined words and the sentence structures for a better workflow, producing the structured patterns of words in a sentence except for some keywords to describe the key findings. 
In addition, there also exist linguistic challenges in medical domain, ranging from the difference in sentence structure like the common usage of the negation to the frequent use of the domain-specific medical terms rarely used in the general domain. For instance, the \textit{"There is no boy in the picture"} would be rather awkward and unlikely to be used in the caption for the photographic image, but the descriptions like \textit{"There is no finding suggesting pneumonia"} are frequently used in radiology reports as it provides important information about the absence of abnormal findings.
These discrepancies indicate the need for the text and image encoders specialized for the medical domain, as well as a novel model architecture that can bridge them.



\begin{figure}[!t]
	\centering
 \centerline{\epsfig{figure=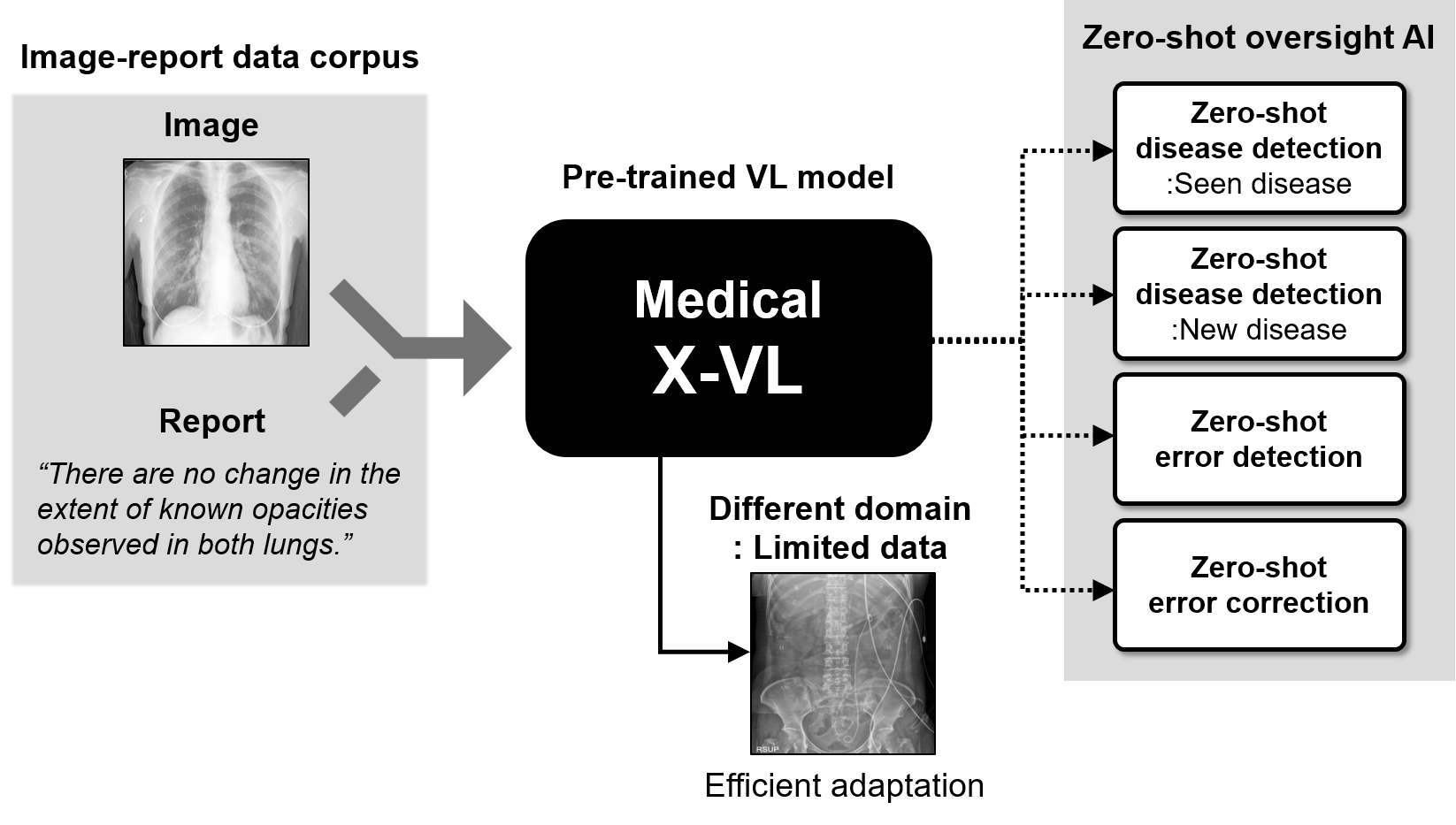, width=0.7\linewidth}}
	\caption{\bf 
The proposed vision-language model, medical X-VL, is self-supervised on an uncurated image-report corpus, and can be used for various zero-shot oversight AI tasks and facilitates efficient adaptation to medical data from different domains.}
\label{fig:app}
\end{figure}

To address this, here we present a model dubbed \textit{Medical Cross-attention Vision-Language model (Medical X-VL)}, leveraging the key components to be tailored for the medical domain.
Our medical X-VL model is based on the following components: self-supervised uni-modal models in medical domain and fusion encoder to bridge them, momentum distillation, sentence-wise contrastive learning for medical reports, and the sentence similarity-adjusted hard negative mining.
We experimentally demonstrated that our model outperforms the current state-of-the-art medical vision-language and self-supervised models in tasks for the oversight AI, ranging from the zero-shot disease detection to detect not only seen disease classes during the pre-training but also unseen disease class like COVID-19 to zero-shot detection and correction of human errors, which can be catastrophic and even life-threatening (\ref{fig:app}).
In addition to the quantitative measures, we performed qualitative analysis on the medical X-VL model by visualizing the cross-attention between images and words, providing another merit to visualize the word-region level attention, providing the transparent interpretation on the model's behavior.
Finally, we extended our model to another medical domain data with limited number, to validate the adaptability of our model in wider clinical applications (Figure~\ref{fig:app}).

\section*{Results}
\subsection{Overview of the proposed model}
Currently, most vision-language models can be classified into two distinct architectures: the parallel dual encoder architecture and the multi-modal fusion encoder architecture. The parallel dual encoder architecture refers to a model that uses cross-modal contrastive loss to allow uni-modal encoders to embed the encoded features in the same embedding space, as suggested in pioneering works in the general domain, such as CLIP\cite{radford2021learning} and ALIGN\cite{jia2021scaling}, as well as in the medical domain\cite{tiu2022expert, zhang2022contrastive, boecking2022making} (Supplementary Fig.~\ref{supple_arch}a).  While this approach is intuitive and capable of positioning vision and text representations in the same embedding space, it relies solely on contrastive loss, which requires a large number of image-text pairs to achieve proper alignment in the embedding space. This can be a significant limitation, particularly in the medical domain, where there may be insufficient image-text pairs compared to the general domain. Additionally, since this approach uses parallel uni-modal encoders without a bridging module between them, there may be limitations in utilizing downstream applications that require multi-modal representations, such as image-guided text generation or completion.
In contrast, the design using a multi-modal fusion encoder performs direct fusion between image and text representations using self-attention (Supplementary Fig.~\ref{supple_arch}b) or cross-attention (X-attention) (Supplementary Fig.~\ref{supple_arch}c). This approach enables obtaining an image-text fused representation, which can be utilized to perform various downstream tasks requiring multi-modal understanding.
Several medical vision-language models have demonstrated the ability to perform diverse tasks requiring multi-modal understanding in the medical domain using this structure\cite{moon2021multi, yan2022clinical}. However, the approach of jointly learning image and text inputs without cross-modal alignment often faces challenges in ensuring that the multi-modal encoder successfully learns image-text interaction, as visual and textual features are not aligned.

A recently proposed method\cite{li2021align} leverages the strengths of both designs by introducing CLIP-style pre-alignment before the multi-modal fusion. Furthermore, to overcome the limitation that the web-scale database used for VLP in the general domain contains many noisy pairs, a momentum distillation method utilizing knowledge distillation from a momentum teacher model was employed to obtain informative features that cannot be obtained through one-hot labels. These innovations make it possible for the model to achieve SOTA performance in various vision-language tasks with relatively few data and enable data-efficient learning of the vision-language model. For more detailed information on 
the concept of the momentum distillation, please refer to the Supplementary material. Building upon prior research and models, we introduce designs that are suitable for medical domain data, specifically medical image and report pairs. Figure~\ref{fig:app} illustrates the proposed medical X-VL model and the learning approach devised for the medical domain.

\begin{figure}[!t]
	\centering
 \centerline{\epsfig{figure=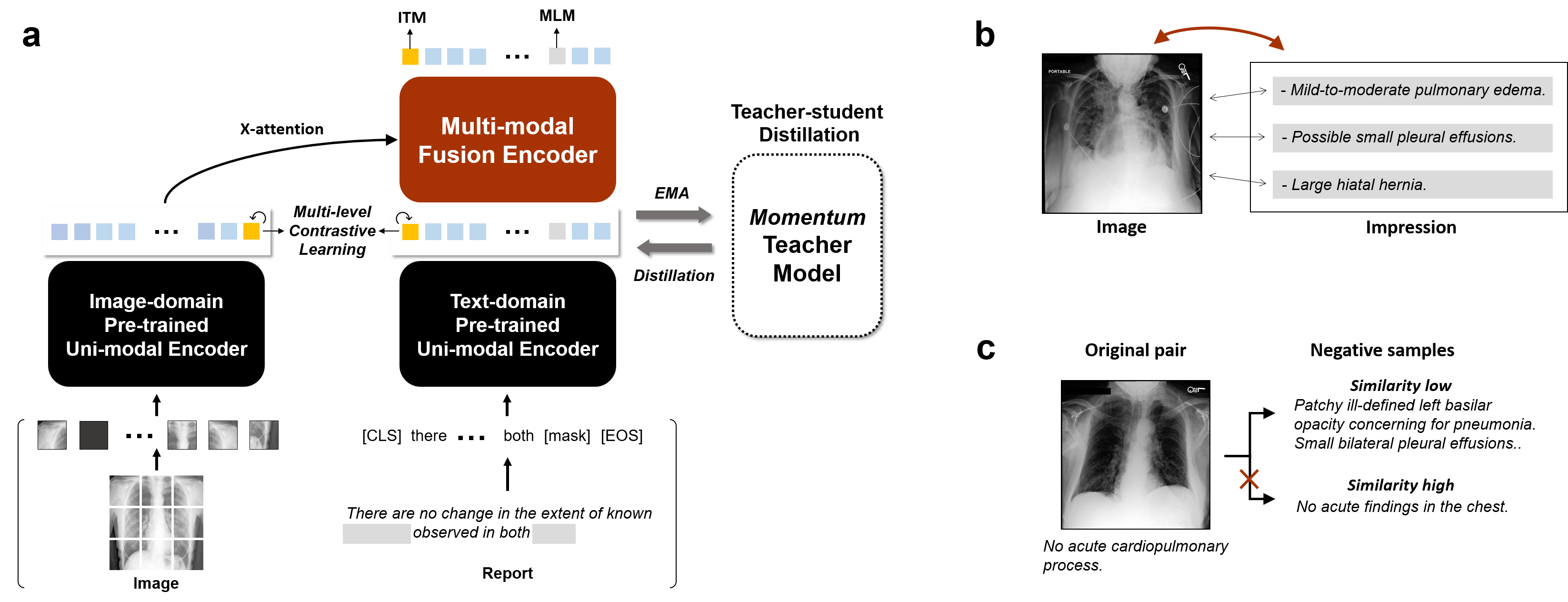, width=1.0\linewidth}}
	\caption{\bf Architecture and key components of the medical X-VL model. (A) The model is based on a cross-attention (X-attention) based multi-modal vision language architecture that utilizes contrastive learning and momentum distillation methods. The cross-attention-based fusion encoder connects the uni-modal encoders that are self-supervised in each uni-modal domain. (B) To account for medical reports being composed of multiple sentences, contrastive learning is introduced between each sentence and image. (C) Since similar reports belonging to the same class are likely to exist between medical reports, text similarity is calculated to exclude those with high similar, thereby improving negative mining performance.
	}
	\label{fig:model}
\end{figure}

Based on the models proposed in previous work\cite{li2021align}, we introduced contraptious methods suitable for medical domain data, specifically medical image and report pairs. Figure \ref{fig:model}a illustrates the proposed learning method consisting of a fusion encoder that performs cross-attention between uni-modal encoders, which are self-supervised model using their respective domain data. As the domain-specific self-supervised model, we used DINO\cite{caron2021emerging} pre-trained model on MIMIC-CXR training set as the visual encoder and the self-supervised CXR-BERT model\cite{boecking2022making} as the text encoder. This approach enables stable training by enabling the fusion module to work mainly to bridge the pre-trained uni-modal encoders during the learning process of vision-language model. 

Additionally, we proposed an additional contrastive learning suitable for medical domain data as shown in Figure \ref{fig:model}b. Unlike a short sentence that describes the overall content of an image in the general domain, a medical report usually consists of several sentences, each representing a specific observation. Inspired by previous studies that used the global-to-local as well as the global-to-global contrastive learnings to improve the performance\cite{yang2022vision}, we introduced contrastive learning between individual sentences and images to ensure that the individual observations can align well with the visual representation. The masked language modeling (MLM) loss enables fine-grained correlation between the image and individual words by predicting the masked words with the help of image information. To compute the image-text matching (ITM) loss that directly determines whether a given image-text pair is correct, it is necessary to construct negative samples. Although hard negative mining proposed in previous works is an effective negative sampling method, it may select a positive sample as a negative sample in medical domain data where multiple images may exhibit the same observation and therefore semantically identical image or report can be sampled as the negative sample in the same batch. Therefore, we alleviate this issue by calculating the text similarity, as shown in Figure \ref{fig:model}c, to avoid selecting those with high text similarities. Ablating any of these elements resulted in sub-optimal performance, as shown in the Supplementary Table~\ref{tab:ablation}, indicating that each component is essential for achieving optimal performance.
For a detailed description of the model architecture and learning objectives, refer to the Method section.

To enable the detection of clinically significant abnormalities and oversight AI for detecting human errors without explicit labeling, we leveraged zero-shot learning. For this purpose, we utilized the MIMIC-CXR dataset, which includes 377,110 image pairs. We trained the model using 371,951 images excluding the test set.

\begin{table}[t!]
  \centering
  \caption{Performance of zero-shot oversight AI to alert clinically significant abnormality}
  \begin{adjustbox}{width=\textwidth}
    \begin{tabular}{lp{7.4em}p{7.4em}p{7.4em}p{7.4em}p{7.4em}p{7.4em}}
    \toprule
    \textbf{} & \multicolumn{1}{l}{\textbf{Average}} & \multicolumn{1}{l}{\textbf{Atelectasis}} & \multicolumn{1}{l}{\textbf{Cardiomegaly}} & \multicolumn{1}{l}{\textbf{Consolidation}} & \multicolumn{1}{l}{\textbf{Pulmonary edema}} & \multicolumn{1}{l}{\textbf{Pleural effusion}} \\
    \midrule
        \textbf{AUC} & \multicolumn{1}{c}{} & \multicolumn{1}{c}{} & \multicolumn{1}{c}{} & \multicolumn{1}{c}{} & \multicolumn{1}{c}{} & \multicolumn{1}{c}{} \\
    X-VL (simple)  & 0.879 \newline{}(0.842-0.911) & 0.803 \newline{}(0.763-0.839) & 0.874 \newline{}(0.841-0.903) & 0.880 \newline{}(0.830-0.923) & 0.903 \newline{}(0.867-0.934) & 0.934 \newline{}(0.910-0.955) \\
    X-VL (detailed)  & 0.881 \newline{}(0.843-0.912) & 0.812 \newline{}(0.773-0.850) & 0.869 \newline{}(0.835-0.899) & 0.897 \newline{}(0.848-0.936) & 0.901 \newline{}(0.863-0.930) & 0.924 \newline{}(0.896-0.948) \\
    CheXzero (simple) & 0.878 \newline{}(0.839-0.913) & 0.787 \newline{}(0.744-0.826) & 0.898 \newline{}(0.870-0.923) & 0.904 \newline{}(0.838-0.955) & 0.888 \newline{}(0.856-0.919) & 0.917 \newline{}(0.888-0.941) \\
    CheXzero (detailed) & 0.830 \newline{}(0.782-0.873) & 0.696 \newline{}(0.647-0.743) & 0.863 \newline{}(0.830-0.894) & 0.809 \newline{}(0.723-0.886) & 0.877 \newline{}(0.840-0.913) & 0.902 \newline{}(0.872-0.930) \\
    \textbf{F1} & \multicolumn{1}{c}{} & \multicolumn{1}{c}{} & \multicolumn{1}{c}{} & \multicolumn{1}{c}{} & \multicolumn{1}{c}{} & \multicolumn{1}{c}{} \\
    X-VL (simple)  & 0.630 \newline{}(0.552-0.705) & 0.666 \newline{}(0.607-0.718) & 0.692 \newline{}(0.636-0.745) & 0.413 \newline{}(0.286-0.552) & 0.638 \newline{}(0.554-0.709) & 0.740 \newline{}(0.677-0.798) \\
    X-VL (detailed)  & 0.629 \newline{}(0.552-0.701) & 0.653 \newline{}(0.595-0.705) & 0.693 \newline{}(0.635-0.747) & 0.433 \newline{}(0.308-0.558) & 0.629 \newline{}(0.548-0.702) & 0.739 \newline{}(0.673-0.796) \\
    CheXzero (simple) & 0.645 \newline{}(0.565-0.720) & 0.641 \newline{}(0.588-0.699) & 0.748 \newline{}(0.692-0.799) & 0.515 \newline{}(0.369-0.657) & 0.612 \newline{}(0.538-0.677) & 0.708 \newline{}(0.638-0.770) \\
    CheXzero (detailed) & 0.579 \newline{}(0.504-0.652) & 0.547 \newline{}(0.489-0.601) & 0.694 \newline{}(0.638-0.745) & 0.351 \newline{}(0.235-0.473) & 0.610 \newline{}(0.530-0.687) & 0.693 \newline{}(0.628-0.754) \\
    Radiologists$^{*}$ & 0.615 \newline{}(0.552-0.672) & 0.692 \newline{}(0.646-0.731) & 0.678 \newline{}(0.634-0.718) & 0.385 \newline{}(0.280-0.485) & 0.583 \newline{}(0.511-0.645) & 0.737 \newline{}(0.689-0.783) \\
    \bottomrule
                    \multicolumn{7}{l}{\footnotesize  * Results are from the previous work\cite{tiu2022expert}.}
    \end{tabular}%
    \end{adjustbox}
  \label{tab:zero_classification_CXR}%
\end{table}%

\subsection{Zero-shot oversight AI to alert clinically significant abnormality}
Detecting clinically significant abnormalities using zero-shot learning can be considered a zero-shot classification problem for the abnormal class. The model's performance was evaluated on the CheXpert competition's test set data of 500 images\cite{irvin2019chexpert}, classifying five abnormality labels, atelectasis, cardiomegaly, consolidation, pulmonary edema, and pleural effusion. Evaluation was conducted in two ways: one was calculating the matching score between the simple prompts of \textit{"{pathology}"} and \textit{"no {pathology}"} as proposed in a previous work\cite{tiu2022expert}, and the other was calculating the score between the \textit{class-specific detailed descriptions} selected by clinicians and \textit{"no {pathology}"} as proposed in another work\cite{zhang2022contrastive}. For example of the detailed description for a given abnormality class, refer to Supplementary Fig.~\ref{supple_detailed}.
Through the latter method, the model's understanding of fine-grained details supporting the class estimation can be evaluated, unlike just simple class classification with simple prompts. We mainly compared our model with the SOTA zero-shot classification model, CheXzero, as well as other medical vision-language models and self-supervised models.

Table~\ref{tab:zero_classification_CXR} shows the model performance compared to the current SOTA model and radiologists. Using the simple prompts, the model showed excellent performance that was not statistically significantly different from the previous SOTA model in terms of area under the receiver operating characteristic curve (AUC), based on the average of metrics for the five classes. Similarly, in terms of F1 score, the model showed better performance based on the average of five classes than the radiologists as well as the the current SOTA model, albeit not statistically significant.
When using detailed description for each abnormality class, the improved performances of the proposed model over the current SOTA model were pronounced. Specifically, in terms of AUC, the proposed model showed statistically significantly better performances for detection of atelectasis, and also showed higher performances in all other labels without statistical significance. In terms of F1-score, the proposed model showed the trend toward better detection performances based on the average of metrics for all abnormality classes than both the current SOTA model and the radiologists.
Compared to the current SOTA model, the proposed model demonstrated trends toward better results in the metric averaged over the five classes. In the zero-shot detection utilizing detailed descriptions for each class, the current SOTA model showed a significant degradation in performance, indicating that it is inadequate for detecting abnormalities based on the detailed descriptions for each class, rather than just the class name. On the other hand, the proposed model exhibited almost no compromise in performance, showing that the model more accurately understands the detailed descriptions that can explain each abnormality class.

Table~\ref{tab:comparison_VLmodels} presents the performance comparison of our proposed model with various medical vision-language models and self-supervised models other than the current SOTA model, CheXzero. The results indicate that our approach achieves better or comparable performance to self-supervised models that have been fine-tuned with some (1-50\%) or all (100\%) labeled data as well as medical vision-language models without the need for explicit label training.

Combined, the results demonstrate that our model can perform zero-shot oversight AI to alert clinically significant abnormalities with a more fine-grained understanding of various abnormalities.

\begin{table}[t!]
  \centering
  \caption{Comparison of zero-shot classification performance with medical vision-language models and self-supervised models}
    \begin{adjustbox}{width=0.5\textwidth}
    \begin{tabular}{rcc}
    \toprule
          & \textbf{Model} & \textbf{Mean AUC} \\
    \midrule
    \multicolumn{1}{l}{\textbf{Supervised}} & DenseNet-121$^{*}$ & 0.901 \\
    \multicolumn{1}{l}{\textbf{Self-supervised}} & GLoRIA$^{*}$ & 0.534 \\
          & ConVIRT-ResNet-50-1\%$^{*}$ & 0.870 \\
          & ConVIRT-ResNet-50-10\%$^{*}$ & 0.881 \\
          & ConVIRT-ResNet-50-100\%$^{*}$ & 0.881 \\
          & ConVIRT-ViT-1\%$^{*}$ & 0.725 \\
          & ConVIRT-ViT-10\%$^{*}$ & 0.809 \\
          & ConVIRT-ViT-100\%$^{*}$ & 0.856 \\
          & CheXzero & 0.878 \\
          & X-VL (ours) & 0.881 \\
    \bottomrule
                \multicolumn{3}{l}{\footnotesize  * Results are from the previous work\cite{tiu2022expert}.}
    \end{tabular}%
        \end{adjustbox}
  \label{tab:comparison_VLmodels}%
\end{table}%

\subsection{Zero-shot oversight AI to detect critical radiology report error}

In radiology, errors can occur in many aspects through the reading process, which lead to fatal results though not frequently occur. 
Based on the result that our model showed the best performance in correlating the detailed components of image and text, we verified whether the trained model can detect critical human errors in a zero-shot manner. Inspired by the previous studies\cite{min2022rred, yu2022evaluating}, we classified human errors into five classes for simulation: image-report mis-registration (mismatch error), error in description for location (location error), error in description for extent (extent error), false estimate of no finding (false-negative error), and false estimate of abnormal finding (false-positive rror). We designed an error generator to produce these errors with a probability of 5\%, and evaluated whether the vision-language model could detect the existence of these errors without any fine-tuning process. For more details on the simulation of the radiology report errors, refer to the Method section.

\begin{table}[t!]
  \centering
  \caption{Zero-shot error detection performance for the critical radiology report errors}
  \begin{adjustbox}{width=\textwidth}
    \begin{tabular}{lp{7em}p{7em}p{7em}p{7em}p{7em}p{7em}}
    \toprule
    \textbf{} & \multicolumn{1}{l}{\textbf{Average}} & \multicolumn{1}{l}{\textbf{Mismatch}} & \multicolumn{1}{l}{\textbf{Location}} & \multicolumn{1}{l}{\textbf{Extent}} & \multicolumn{1}{l}{\textbf{False-negative}} & \multicolumn{1}{l}{\textbf{False-positive}} \\
    \midrule
        \textbf{AUC} & \multicolumn{1}{c}{} & \multicolumn{1}{c}{} & \multicolumn{1}{c}{} & \multicolumn{1}{c}{} & \multicolumn{1}{c}{} & \multicolumn{1}{c}{} \\
    X-VL  & 0.855 \newline{}(0.796-0.905) & 0.956 \newline{}(0.937-0.972) & 0.724 \newline{}(0.640-0.797) & 0.703 \newline{}(0.602-0.804) & 0.926 \newline{}(0.890-0.958) & 0.966 \newline{}(0.913-0.995) \\
    CheXzero & 0.744 \newline{}(0.686-0.799) & 0.826 \newline{}(0.797-0.854) & 0.589 \newline{}(0.516-0.666) & 0.638 \newline{}(0.552-0.717) & 0.829 \newline{}(0.798-0.865) & 0.839 \newline{}(0.765-0.895) \\
    \textbf{F1} & \multicolumn{1}{c}{} & \multicolumn{1}{c}{} & \multicolumn{1}{c}{} & \multicolumn{1}{c}{} & \multicolumn{1}{c}{} & \multicolumn{1}{c}{} \\
    X-VL  & 0.516 \newline{}(0.392-0.631) & 0.621 \newline{}(0.526-0.691) & 0.315 \newline{}(0.206-0.431) & 0.352 \newline{}(0.215-0.500) & 0.623 \newline{}(0.522-0.710) & 0.672 \newline{}(0.493-0.821) \\
    CheXzero & 0.171 \newline{}(0.126-0.227) & 0.288 \newline{}(0.237-0.339) & 0.098 \newline{}(0.063-0.146) & 0.113 \newline{}(0.073-0.156) & 0.238 \newline{}(0.188-0.292) & 0.119 \newline{}(0.068-0.200) \\
    \bottomrule
    \end{tabular}%
    \end{adjustbox}
  \label{tab:zeroshot_error_CXR}%
\end{table}%

Overall, our model significantly outperformed the current SOTA model in the oversight task of detecting human errors (Table~\ref{tab:zeroshot_error_CXR}). When examining each type of error, our proposed model provided significantly better performances for the mismatch, false-negative and false-positive errors in terms of AUC, although the performances for the other error classes were also better without statistical significance. In terms of F1-score, the proposed model showed better detection performances for all error types with statistical significances.
Considering that detecting errors in sentences requires fine-grained understanding of the components of the sentence than simply classifying image with the corresponding class, the result again demonstrates the superiority of our model in terms of detailed understanding of sentence components.

\begin{figure}[!t]
	\centering
 \centerline{\epsfig{figure=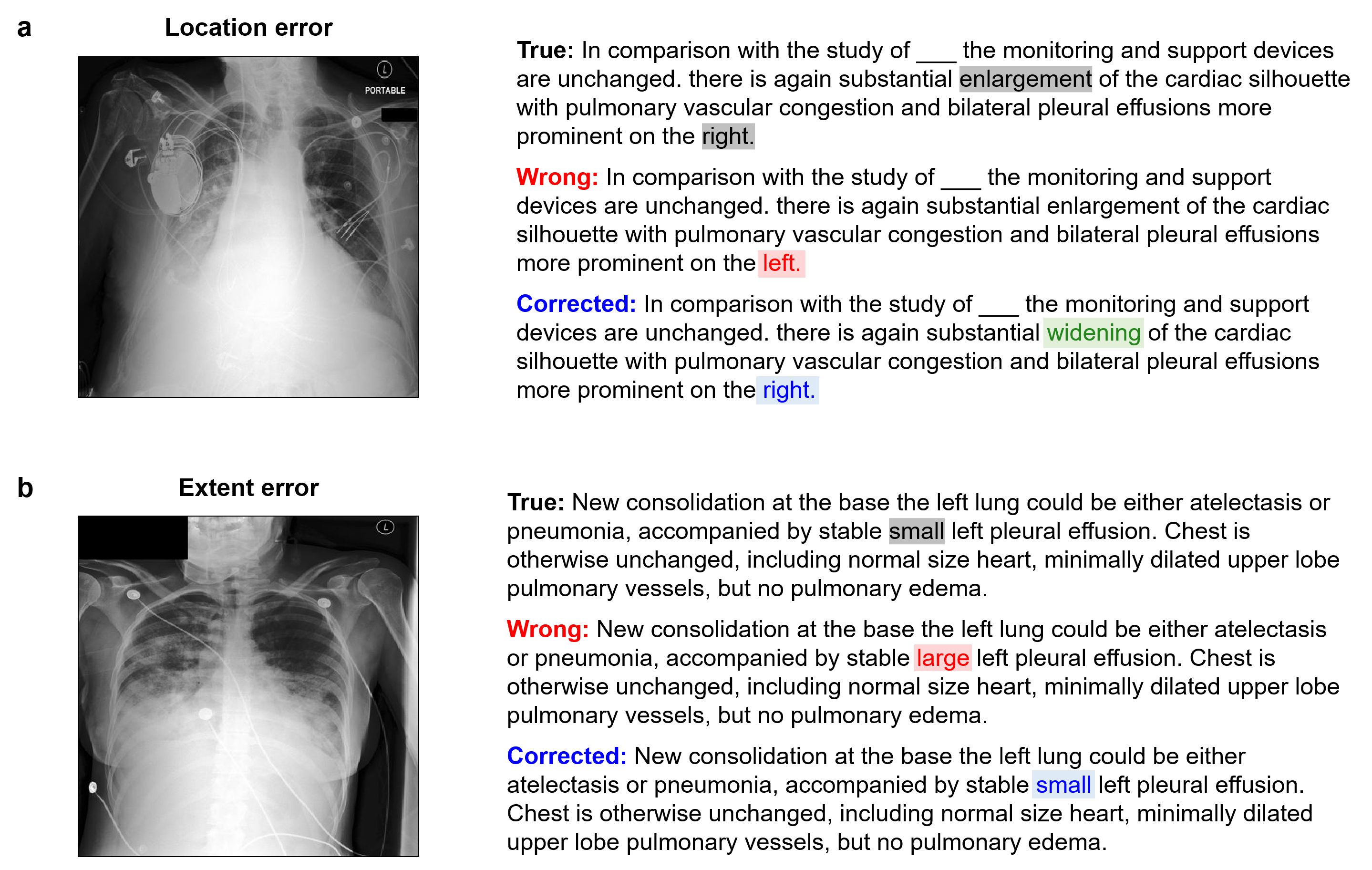, width=1.0\linewidth}}
	\caption{\bf 
(A) Example for the correction of the location error. Notably, other than the wrong words ("left" $\rightarrow$ "right"), it also changes another word ("enlargement") to another ("widening") without significantly changing the meaning.
(B) Example for the correction of the extent error ("large" $\rightarrow$ "small").}
	\label{fig:correction}
\end{figure}

The structure of the medical X-VL model with a multi-modal fusion module provides another merit of enabling zero-shot "correction" as well as zero-shot detection of errors. Thanks to the image-guided MLM in pre-training objectives, the model can correct erroneous expression by substituting wrong words (red text) with correct ones (blue text) referring to the image when masking each word within a sentence one by one and predicting it (Figure~\ref{fig:correction}). The model was able to replace not only the words with "location error (Figure~\ref{fig:correction}a)" but also words with "extent error (Figure~\ref{fig:correction}a)" with those of similar meanings. Interestingly, it was observed that even another word (e.g. enlargement) without error was replaced with different words with similar meanings (e.g. widening). This behavior of the model was understandable, considering the nature of one-by-one substitution approach using MLM by the model.




\subsection{Model applicability to unseen disease and scalability of uni-modal pre-trained model to multi-modal understanding}

The MIMIC-CXR dataset consists of patient data collected from 2011 to 2016 and does not include cases of coronavirus disease (COVID-19) that first emerged in late 2019. Therefore, for models trained with vision-language pre-training using MIMIC-CXR, COVID-19 can be considered an unseen disease. The detailed findings (bilateral peripheral and basal multifocal airspace ground glass opacity or consolidation) indicating COVID-19 can also be observed in other infectious diseases, suggesting that accurate diagnosis of new diseases can be achieved with high accuracy by detecting the presence of these detailed findings. Table~\ref{tab:unseen} shows zero-shot detection performances of unseen abnormality of COVID-19.
When using the detailed description for zero-shot detection, there was no statistically significant difference between the proposed model and the current SOTA model, although the proposed model showed trend toward better performances for both AUC and F1-score.

The term \textit{"COVID-19"} is not included in the training of vision-language models, and therefore, direct detection using a simple prompt with \textit{"{COVID-19}"} and \textit{"no {COVID-19}"} is expected to result in significant performance degradation. Interestingly, while the current state-of-the-art model shows a marked deterioration, our proposed model does not experience a decrease in performance, presenting significantly better performance in terms of both AUC and F1-score.
The success of our model can be attributed to its design, which bridges domain-specific pre-training models. The text encoder of our model employs CXR-BERT\cite{boecking2022making}, a self-supervised model trained not only on MIMIC-CXR and MIMIC-III data, but also on PubMed abstracts, which contain COVID-19 information from recent publications. As a result, the in-domain knowledge acquired during uni-modal self-supervised learning on a text corpus was extended to the multi-modal domain, facilitating the effective detection of previously unseen diseases. These results suggest the potential to expand the knowledge of a uni-modal model pre-trained in a specific domain to other domains through multi-modal combination, highlighting the scalability of currently available large uni-modal pre-trained models.

\begin{table}[t!]
  \centering
  \caption{Zero-shot detection performance for unseen abnormality of COVID-19}
    \begin{adjustbox}{width=0.48\textwidth}
    \begin{tabular}{lp{8.625em}p{9.815em}}
    \toprule
          & \multicolumn{1}{c}{\textbf{Direct class name}} & \multicolumn{1}{c}{\textbf{Detailed impression}} \\
      \midrule
    \textbf{AUC} & \multicolumn{1}{c}{} & \multicolumn{1}{c}{} \\
    X-VL  & 0.799 \newline{}(0.781-0.816) & 0.800 \newline{}(0.785-0.816) \\
    CheXzero & 0.684 \newline{}(0.664-0.702) & 0.778 \newline{}(0.760-0.795) \\
    \textbf{F1} & \multicolumn{1}{c}{} & \multicolumn{1}{c}{} \\
    X-VL & 0.823 \newline{}(0.809-0.836) & 0.802 \newline{}(0.788-0.815) \\
        CheXzero  & 0.722 \newline{}(0.707-0.737) & 0.798 \newline{}(0.784-0.811) \\
    \bottomrule
    \end{tabular}%
    \end{adjustbox}
  \label{tab:unseen}%
\end{table}%

\begin{figure}[!t]
	\centering
 \centerline{\epsfig{figure=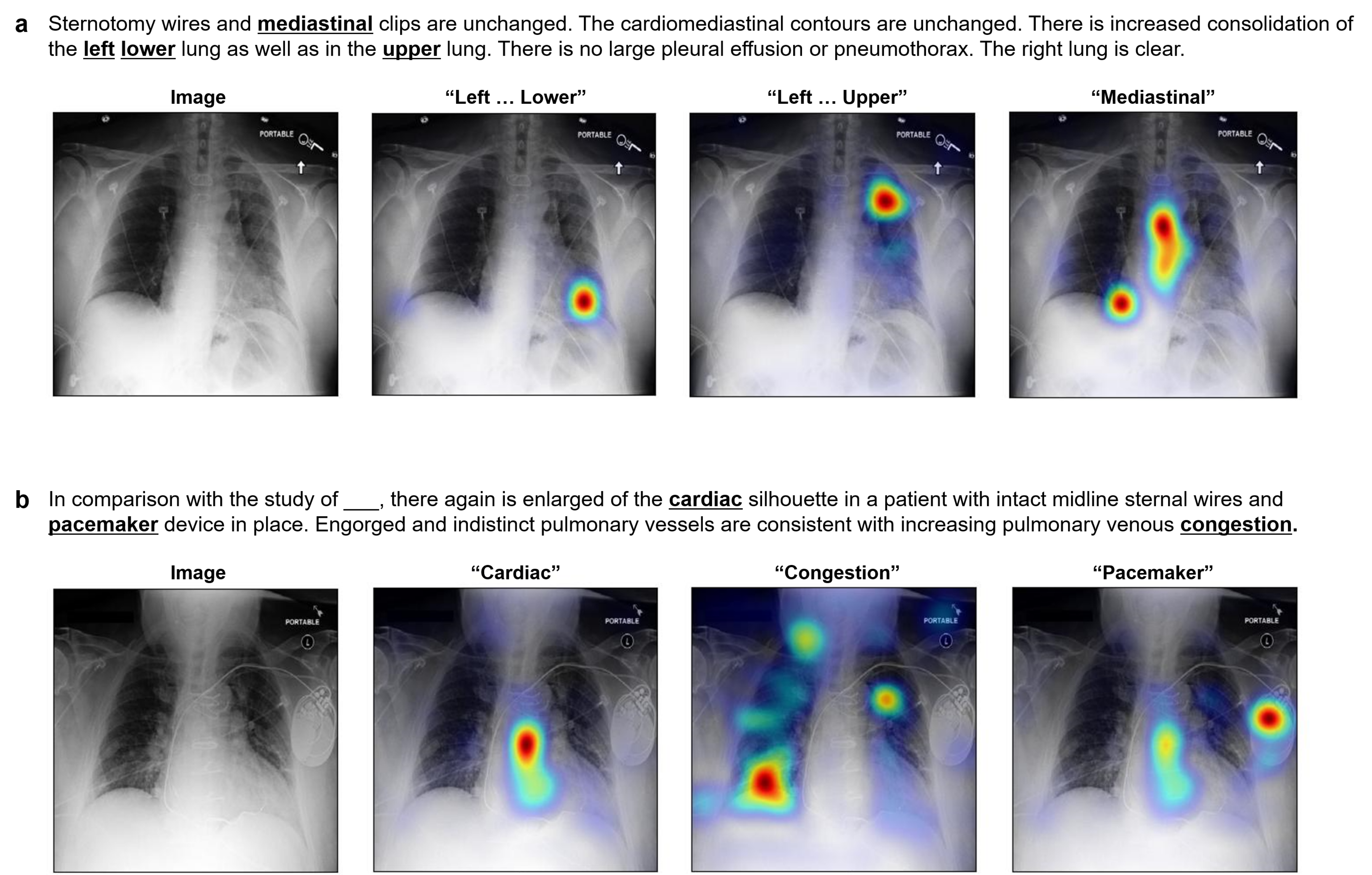, width=1.0\linewidth}}
	\caption{\bf 
(A-B) Exemplified cases of the Grad-CAM visualization of the cross-attention maps corresponding to each word. The medical X-VL model not only grounds the important clinical findings (``Congestion", ``Pacemaker") but also understands the location (``Mediastinal", ``Cardiac") and the relationships (``Left ... Lower", ``Left ... Upper").}
	\label{fig:attention}
\end{figure}

\subsection{Verification of image-text correlation via qualitative analysis of cross-attention}
In contrast to CLIP-based models that visualize task-agnostic self-attention or only the class-level attention, our cross-attention-based model offers the additional advantage of visualizing word-patch level cross-attention between images and sentences. This provides a more transparent interpretation of the model's behavior by considering the meaning of each word component. To achieve this, we performed qualitative analysis using Grad-CAM\cite{selvaraju2017grad} visualization for the fusion module's cross-attention, as suggested in ALBEF\cite{li2021align}, as shown in the illustrated cases of Fig.~\ref{fig:attention}. Without any supervision for the region-word correlations, the medical X-VL model correctly focuses on the regions related to each word, demonstrating its ability to understand the relationship between the semantics of the image and the textual concept. Notably, the model not only identifies important clinical findings such as "Congestion" and "Pacemaker," but also understands the location ("Mediastinal," "Cardiac") and relationships ("Left ... Lower," "Left ... Upper").

\subsection{Application to clinical data in different domain}
In order to further demonstrate the broad applicability of our model, we extended our research to the field of abdominal radiography. We obtained a total of 5,772 image-text pairs of abdominal radiographs from Chung Ang University Hospital (CAUH) for training, and an additional 734 abdominal radiographs from the Chungnam National University Hospital (CNUH) registry for external validation. The model was initialized with pre-trained weights from chest radiographs, as there are similarities between chest and abdominal radiographs, despite differences in the field of view. The main abnormal findings in abdominal radiographs, ileus and pneumoperitoneum, were evaluated. Unlike chest radiographs, the medical reports for abdominal radiographs typically do not include descriptions of the location or extent of the disease, but rather simplified reports. Therefore, the zero-shot error detection performance was evaluated only for three types of errors: mismatch, false-negative, and false-positive.

In evaluating the trained model for zero-shot detection of major abnormal findings in abdominal radiograph, the proposed model showed trend toward better performance than the current SOTA model both with the simple prompt and the detailed prompt for all abnormality classes, although not statistically significant, in terms of both AUC and F1-score (Table~\ref{tab:zeroshot_classification_abdomen}). The trend toward better performance of the proposed model was more pronounced in zero-shot detection of the error showing the larger differences in performance in terms of AUC for all error classes (Table~\ref{tab:zeroshot_error_abdomen}), demonstrating once again the superiority of our proposed method for zero-shot oversight AI in different domain.

\begin{table}[t]
  \centering
  \caption{Performance of zero-shot oversight AI to alert clinically significant abnormality for abdominal radiograph}
      \begin{adjustbox}{width=0.70\textwidth}
    \begin{tabular}{lp{9em}p{9em}p{9em}}
    \toprule
    \textbf{} & \multicolumn{1}{l}{\textbf{Average}} & \multicolumn{1}{l}{\textbf{Ileus}} & \multicolumn{1}{l}{\textbf{Pneumoperitoneum}} \\
    \midrule
        \textbf{AUC} & \multicolumn{1}{c}{} & \multicolumn{1}{c}{} & \multicolumn{1}{c}{} \\
    X-VL (simple)  & 0.837 \newline{}(0.765-0.898) & 0.851 \newline{}(0.810-0.887) & 0.824 \newline{}(0.720-0.910) \\
    X-VL (detailed) & 0.827 \newline{}(0.755-0.890) & 0.837 \newline{}(0.796-0.875) & 0.817 \newline{}(0.714-0.905) \\
        CheXzero (simple)  & 0.767 \newline{}(0.696-0.833) & 0.770 \newline{}(0.716-0.823) & 0.765 \newline{}(0.676-0.843) \\
    CheXzero (detailed) & 0.815 \newline{}(0.749-0.875) & 0.827 \newline{}(0.781-0.872) & 0.803 \newline{}(0.717-0.879) \\
    \textbf{F1} & \multicolumn{1}{c}{} & \multicolumn{1}{c}{} & \multicolumn{1}{c}{} \\
    X-VL (simple)  & 0.441 \newline{}(0.321-0.556) & 0.449 \newline{}(0.376-0.520) & 0.432 \newline{}(0.267-0.592) \\
    X-VL (detailed) & 0.447 \newline{}(0.316-0.571) & 0.420 \newline{}(0.346-0.493) & 0.474 \newline{}(0.286-0.650) \\
        CheXzero (simple)  & 0.317 \newline{}(0.221-0.428) & 0.372 \newline{}(0.300-0.457) & 0.262 \newline{}(0.143-0.400) \\
    CheXzero (detailed) & 0.366 \newline{}(0.276-0.463) & 0.454 \newline{}(0.372-0.540) & 0.279 \newline{}(0.180-0.386) \\
    \bottomrule
    \end{tabular}%
    \end{adjustbox}
  \label{tab:zeroshot_classification_abdomen}%
\end{table}%

\begin{table}[htbp]
  \centering
  \caption{Zero-shot error detection performance for the critical radiology report errors in abdominal radiograph}
        \begin{adjustbox}{width=0.8\textwidth}
    \begin{tabular}{lp{9em}p{9em}p{9em}p{9em}}
    \toprule
    \textbf{} & \multicolumn{1}{l}{\textbf{Average}} & \multicolumn{1}{l}{\textbf{Mismatch}} & \multicolumn{1}{l}{\textbf{False-negative}} & \multicolumn{1}{l}{\textbf{False-positive}} \\
    \midrule
        \textbf{AUC} & \multicolumn{1}{c}{} & \multicolumn{1}{c}{} & \multicolumn{1}{c}{} & \multicolumn{1}{c}{} \\
    X-VL  & 0.816 \newline{}(0.717-0.900) & 0.837 \newline{}(0.769-0.892) & 0.785 \newline{}(0.627-0.929) & 0.826 \newline{}(0.756-0.881) \\
    CheXzero & 0.616 \newline{}(0.414-0.811) & 0.695 \newline{}(0.558-0.807) & 0.426 \newline{}(0.105-0.778) & 0.728 \newline{}(0.580-0.849) \\
    \textbf{F1} & \multicolumn{1}{c}{} & \multicolumn{1}{c}{} & \multicolumn{1}{c}{} & \multicolumn{1}{c}{} \\
    X-VL  & 0.263 \newline{}(0.151-0.408) & 0.335 \newline{}(0.209-0.451) & 0.102 \newline{}(0.008-0.303) & 0.353 \newline{}(0.236-0.469) \\
    CheXzero & 0.230 \newline{}(0.129-0.331) & 0.327 \newline{}(0.188-0.443) & 0.030 \newline{}(0.003-0.083) & 0.334 \newline{}(0.196-0.467) \\
    \bottomrule
    \end{tabular}%
        \end{adjustbox}
  \label{tab:zeroshot_error_abdomen}%
\end{table}%

\section*{Discussion}

Despite the remarkable achievements of deep learning-based AI models in various tasks, their successes have been limited to narrow domains, suggesting the presence of fundamental challenges yet to be addressed.
While contemporary AI-based CAD models exhibit outstanding performance in detecting abnormalities within an image, they are primarily designed as standalone diagnostic tools that often disrupt common radiological workflows. Moreover, they lack the complementary ability to assist human readers in identifying errors in image descriptions due to their inability to interpret both image and text jointly.


Unlike the model trained with traditional supervised learning, where image-label pairs are manually annotated to train visual recognition, vision-language model uses uncurated image-text pairs. This approach has the advantage of allowing the model to learn rich semantics and broad coverage of visual concepts from free-form text, rather than being confined to discrete labels that offer dense but limited visual concepts\cite{yang2022unified}. By learning broad visual semantics and corresponding textual concepts together, the model can achieve good performance across a range of downstream tasks. However, a limitation of this approach is that the image-text pair lacks the powerful discriminative ability of dense labels used in traditional supervised learning. As a result, training a robust vision-language model often requires billion-scale image-text data, which is difficult to obtain in the medical domain, as demonstrated in the studies of CLIP\cite{radford2021learning} and ALIGN\cite{jia2021scaling}.

To overcome these issues, we utilized several key components in our study. Specifically, we adopted a multi-modal fusion encoder that bridges domain-specific pre-trained uni-modal models, leveraging domain-specific self-superivsed learning with single domain data and thereby allowing for stable learning even with fewer image-text pairs. We also employed momentum distillation, which was originally developed and used in a previous work\cite{li2021align}, to effectively learn information that cannot be learned from one-hot labels and to alleviate the problems of noisy data pairs. This method was also found to be effective in medical domain data, which tends to have relatively small difference even between negative pairs, both visually and linguistically, than general domain data. There may exist weak positive relations with partially overlapping impressions even between negative pairs, and thus, a penalization method different to the one-hot approach is demanded. For this purpose, momentum distillation can be utilized, as the momentum teacher can provide different pseudolabels for pairs with different similarities, allowing to learn additional information between the labels. 
Additionally, we designed a sentence-wise contrastive learning method between individual sentences that comprise a medical report and images, taking into account that medical reports are composed of multiple sentences, unlike general domain text captions with single sentence. This method is an extension of the global-to-local contrastive learning method\cite{yang2022vision} and allows images to be more closely correlated with the semantic components of individual sentences.

As a result, this structure enabled the model to better understand fine-grained image-sentence relations in the medical domain, allowing it to successfully perform not only zero-shot abnormality detection, but also error detection. The performance difference between our model and the current SOTA models was more pronounced in error detection, where the model should also detect partial errors of the given sentence, as well as when using a detailed prompt rather than a simple prompt. This implies that the model has a better understanding of more detailed meaning of words that constituting a sentence, beside the simple class names to produce more accurate results.
Moreover, our proposed method offers several additional benefits. In contrast to previous models that employ separate image and text encoders for direct alignment\cite{tiu2022expert, zhang2022contrastive, boecking2022making}, our approach utilizes a fusion encoder with multi-modal cross attention to facilitate image-guided text prediction through the MLM objective in pre-training. This allows for zero-shot error correction of incorrect words.
The fusion of uni-modal self-supervised models through a fusion encoder may further enhance the model's scalability by leveraging large amounts of single-domain data with self-supervised learning. Our experiments with the unseen COVID-19 disease class demonstrated that extending the concept of "COVID-19" from the uni-modal self-supervised text encoder to the multi-modal domain was achievable. Given that well-aligned data in the medical domain is relatively scarce while large-scale domain-specific data is more abundant, this property suggests the potential to improve model scalability with single-domain data, reducing reliance on well-aligned image-text pairs.

Our study is not free of limitations. Firstly, despite our efforts to reduce reliance on labeled image-text data through domain-specific self-supervised learning, our approach still exhibits a certain degree of dependency on image-text paired data. Secondly, our zero-shot correction method using single word masking is effective only for correcting single-word errors and may not be suitable for correcting errors involving sentences of varying lengths, such as mis-registrations. Lastly, although we have developed a cross-attention-based model that demonstrates improved fine-grained understanding of semantics compared to previous approaches, the detection performance for location and extent errors, which involve subtle changes in a few words, was lower compared to other types of errors.

Despite the limitations, our proposed medical X-VL model, specifically designed for the medical domain, has demonstrated remarkable performance in various oversight AI tasks using a zero-shot approach, surpassing the current SOTA models. As AI models are increasingly used to assist medical professionals rather than replace them, our model trained on uncurated data without explicit labeling, but with a flexible understanding of vision-language concepts, would be highly beneficial in various clinical applications. Furthermore, the image-report pair structure and the multiple-sentence report structure are commonly found in medical imaging modalities beyond radiographs, suggesting that our approach has broad applicability in the field of medical imaging.

\section*{Methods}
\subsection{Details of model architecture}
Our medical X-VL model is based on the commonly used cross-attention-based multi-modal vision-language models such as those introduced in recent studies\cite{li2021align, yang2022vision, dou2022coarse, yu2022coca}. It connects the pre-trained uni-modal models in the medical domain by utilizing the CXR-BERT model\cite{boecking2022making} as the text encoder, which is a transformer model consisting of 12 layers and 12 multi-heads, and the vision transformer pre-trained on MIMIC training data using DINO self-supervised learning\cite{caron2021emerging} for the vision encoder.

As the fusion encoder, we opted for the BERT${base}$ model\cite{devlin2018bert}. The image encoder transforms the input image $I$ into a sequence of patch embeddings {$p_{cls}, p_1, ..., p_N$}, where $p_{cls}$ denotes the \texttt{[CLS]} token embedding. Similarly, the input text $T$ is converted into a sequence of word embeddings {$w_{cls}, w_1, ..., w_M$}, where $w_{cls}$ represents the \texttt{[CLS]} token indicating the start of the sequence, and $w_M$ signifies the end of the sentence as the \texttt{[SEP]} token. By utilizing cross-attention between the modalities, the word embeddings and patch embeddings are fused in the multi-modal fusion encoder, producing the fused word embeddings {$v_{cls}, v_1, ..., v_M$}.


\subsection{Details of pre-training objectives}
The medical X-VL model is trained using three distinct learning objectives, namely contrastive learning for achieving cross- and intra-modal alignment, masked language modeling (MLM) to facilitate image-guided text completion, and image-text matching (ITM).


\paragraph{\small Contrastive Learning for Cross- and Intra-modal Alignment}
The objective of cross-modal contrastive learning is to align image and text features in a shared embedding space prior to fusion, through the application of uni-modal encoders. It operates by attracting positive image-text pairs towards each other, while pushing unmatched pairs apart.
In the context of the encoded embeddings $p_{cls}$ and $w_{cls}$ of the \texttt{[CLS]} tokens of image $I$ and text $T$, respectively, the similarity functions $sim(I, T)$ and $sim(T, I)$ can be defined as follows:

\begin{linenomath}
\begin{align}
\label{Eq1}
sim(I, T) = h_I(p_{cls})^\top h_T(w_{cls}), \quad sim(T, I) = h_t(w_{cls})^\top h_I(p_{cls})
\end{align}
\end{linenomath}
where $h_I$ and $h_T$ denote linear projectors with normalization layers for image and text features.
Then, the normalized image-to-text and text-to-image similarities of each image-text pair are calculated as:
\begin{linenomath}
\begin{align}
\label{Eq2}
s_{i2t} = {{\exp{({sim(I, T_m)}/{\tau})}\over{\Sigma_{m=1}^{M}\exp{({sim(I, T_m)}/{\tau})}}}}, \quad s_{t2i} = {{\exp{({sim(T, I_n)}/{\tau})}\over{\Sigma_{n=1}^{N}\exp{({sim(T, I_n)}/{\tau})}}}}
\end{align}
\end{linenomath}
where $\tau$ denotes the temperature parameter.

In contrast, intra-modal contrastive learning focuses on training the model to distinguish positive and negative samples within each modality based on their semantic differences. The normalized similarities between images and between texts can be defined similarly to those in cross-modal contrastive learning.

\begin{linenomath}
\begin{align}
\label{Eq3}
s_{i2i} = {{\exp{({sim(I, I_n)}/{\tau})}\over{\Sigma_{n=1}^{N}\exp{({sim(I, I_n)}/{\tau})}}}}, \quad s_{t2t} = {{\exp{({sim(T, T_m)}/{\tau})}\over{\Sigma_{m=1}^{M}\exp{({sim(T, T_m)}/{\tau})}}}}
\end{align}
\end{linenomath}
where $\tau$ is the same temperature parameter used in the cross-modal contrastive learning.

Consequently, given the one-hot label similarity $y$, the cross-modal contrastive loss $L_{CMC}$, intra-modal contrastive loss $L_{IMC}$ and overall contrastive loss $L_{contrastive}$ can be defined as the cross-entropy loss $H$:

\begin{linenomath}
\begin{align}
\label{Eq4}
L_{CMC} = {1\over2}[H(y_{i2t}, s_{i2t}) + H(y_{t2i}, s_{t2i})]
\end{align}
\end{linenomath}
\begin{linenomath}
\begin{align}
\label{Eq5}
L_{IMC} = {1\over2}[H(y_{i2t}, s_{i2t}) + H(y_{t2i}, s_{t2i})]
\end{align}
\end{linenomath}
\begin{linenomath}
\begin{align}
\label{Eq6}
L_{contrastive} = L_{CMC} + L_{IMC}
\end{align}
\end{linenomath}



In our study, we employed the recent contrastive learning techniques\cite{he2020momentum} and utilized image and text queues to store the most recent $Q$ samples from the momentum encoder for each modality. The size of the queue was set to 40,920 in our experiments. By calculating feature similarities between image and text with the aforementioned objectives, we applied hard negative mining for ITM, where negative pairs with high similarity were sampled more frequently.


\subsection{Masked Language Modeling for Image-guided Text Completion}
The Masked Language Modeling (MLM) learning objective is commonly used to obtain language comprehension in which the model predicts the correct ground truth for masked word tokens $w_{mask}$ while using the corresponding image as a guide. The masking process randomly replaces 15\% of the word tokens with the \texttt{[MASK]} token, with 80\% probability, a random word token with 10\% probability, and the original token with 10\% probability\cite{devlin2018bert}. The masked text is represented as $T^{mask}$, the fusion encoder's prediction for \texttt{[MASK]} tokens is represented as $p^{mask}(I, T^{mask})$, and the ground truth for each word token is represented as $y^{mask}_{w}$. The MLM loss is defined using the cross-entropy loss $H$ in our model.

\begin{linenomath}
\begin{align}
\label{Eq7}
L_{MLM} = H(y^{mask}_{w}, p^{mask}(I, T^{mask}))
\end{align}
\end{linenomath}

The MLM task in our model, where the predictions for the \texttt{[MASK]} token are generated by the multi-modal fusion encoder that incorporates image representation into text tokens, can be seen as an image-assisted prediction of masked text tokens. This allows the model to learn the joint representation of images and text and their interdependence.

\subsection{Image-Text Matching}
The image-text matching task aims to determine whether a given image-text pair is matched or unmatched. We used the fusion embeddings of two \texttt{[CLS]} tokens, which are obtained from the outputs of the image-to-text and text-to-image paths of the fusion encoder. These embeddings reflect the joint representation of the image-text pair. Binary classifiers were added after the fusion embeddings to predict whether an image-text pair is matched or unmatched, resulting in the prediction $c^{itm}(I, T)$. The ITM loss was defined using the cross-entropy loss $H$, where $y^{itm}$ represents the ground truth label for image-text matching, as shown below:

\begin{linenomath}
\begin{align}
\label{Eq9}
L_{ITM} = H(y^{itm}, c^{itm}(I, T))
\end{align}
\end{linenomath}

We also implemented hard negative mining in the ITM task by prioritizing samples with high similarity scores $s_{i2t}$ and $s_{t2i}$ when selecting negative pairs from the batch for a given image or text\cite{li2021align}. This approach enables the model to better distinguish between semantically similar but fine-grained different images, which is particularly important in medical imaging, where images tend to have small differences due to standardized acquisition protocols. By using this strategy, we achieved performance improvements without any additional computational cost.

Combined, the overall pre-training objective $L$ of the medical X-VL is:
\begin{linenomath}
\begin{align}
\label{Eq10}
L = L_{contrastive} + L_{MLM} + L_{ITM}
\end{align}
\end{linenomath}

However, medical images can belong to multiple classes, and different images may represent the same clinical findings. For instance, a pair of image-report with "No specific finding" and another pair with "No acute cardiopulmonary process" are different but describe the same clinical findings. Therefore, if we use the sample of "No acute cardiopulmonary process" as a negative sample for "No specific finding," which actually corresponds to a positive sample, an incorrect supervisory signal would be given, resulting in label noise that can negatively affect the learning process. Hence, we obtained the text feature of a sample obtained by hard negative mining and computed the cosine similarity between this text feature and that of the positive sample. If the cosine similarity exceeded the pre-defined threshold, the negative sample was resampled to ensure the cosine similarity was below this threshold.

\subsection{Momentum Distillation}
In contrastive learning, negative samples may have similar context to positive ones and should be treated differently from entirely different negative samples. For example, when a radiograph shows bibasilar opacifications suggesting pneumonia, a report like "There exist lung opacities in both lower lobe suggesting the severe pleural effusion," which describes a different etiology for opacification, may be regarded as a negative sample. However, it should be penalized differently from an entirely unmatching description like "Both lung fields are clear and there is no remarkable finding." This issue is more critical in the medical domain, where the differences between images and reports are smaller than those between photographic images and captions, and the overlapping between the images or reports can be substantial. Similarly, for MLM, there may be other candidates with semantically identical meaning to the ground truth, like "No remarkable findings" and "No abnormality." However, the binary and one-hot coded labels for contrastive learning and MLM penalize all negative samples without considering their correctness.


To address this problem, we utilized momentum teachers that generate pseudo labels for contrastive learning and MLM. These teachers are gradually updated with exponential moving averaging of the current models. During training, the model aims to match the pseudo labels generated from the momentum teacher by minimizing the distillation loss $L_{dist}$ in addition to the overall loss $L$, with a weight parameter $\lambda$ to balance the contributions of momentum distillation. This approach enables the model to better handle negative samples with similar context and achieve improved performance.

\begin{linenomath}
\begin{align}
\label{Eq11}
L_{total} = (1 - \lambda) \cdot L + \lambda \cdot L_{dist}
\end{align}
\end{linenomath}



\subsection{Details of dataset for vision-language model training}
To traing the vision-language model using chest radiographs, we employed the MIMIC-CXR dataset\cite{johnson2019mimic}, which is an open-source database containing 377,110 pairs of images and their corresponding free-text radiology reports from 227,835 radiographic studies. The full radiology report is made up of various sections, including examination, indication, impression, technique, and comparison. As the impression section contains important summaries of the findings for the image, we used it for the training of vision-language model. We excluded 5,159 images without impressions among the images, used 371,951 images for training based on the official split, and used 3,651 images for zero-shot error detection evaluation on the test set.

To apply our method to other domains with limited data, we utilized the inpatient abdominal radiograph databases of two hospitals for training and evaluation. Specifically, we employed 5,772 images from CAUH for training and collected 734 images from the database of CNUH for external validation.

\subsection{Details of the simulation for human error in radiology}
Radiology reports generated by clinicians can be subject to errors in various ways. To categorize clinically significant errors that occur, our study identified five types: mis-registration, incorrect location, incorrect extent, false-positive estimate, and false-negative estimate. To implement these errors, we designed an error generator to automatically produce them with a given probability. 

When reports in the MIMIC-CXR dataset are classified using the CheXpert labeler, they can be divided into "no finding" and positive and negative estimates for each "abnormality" class. 
Mis-registration errors occur when a report from a different label class is erroneously registered (Supplementary Fig.~\ref{supple_report}a). This can occur when a previous patient's report is attached to the next patient, leading to potential fatal scenarios in clinical practice. To simulate this, we probabilistically replace the original report with a report from another label class. 
Incorrect location errors refer to errors where the description of a lesion's location (left-right, upper-lower, apical-basal, central-peripheral) is incorrect (Supplementary Fig.~\ref{supple_report}b). This can lead to confusion in making clinical diagnoses and treatment decisions. We implement this error type by replacing location-descriptive terms with their opposites (e.g., $right \rightarrow left$). 
Incorrect extent (severity) errors occur when the extent or severity of the findings (mild-severe, small-large, or minimal-extensive) is inaccurately described (Supplementary Fig.~\ref{supple_report}c). This can result in erroneous treatment plans due to the inaccurate information about disease severity. This error type is implemented by substituting words describing extent or severity with their antonyms (e.g., $mild \rightarrow severe$). 
False-positive estimate errors occur when there is actually no abnormal finding, but it is incorrectly interpreted as having some abnormality (Supplementary Fig.~\ref{supple_report}d). This can result in serious problems such as unnecessary treatment. We implement this error type by first determining if the report corresponds to "no finding," and then replacing it with a report corresponding to a certain abnormality. 
False-negative estimate errors occur when there is an abnormality present but not described in the report and is regarded as no finding (Supplementary Fig.~\ref{supple_report}e). This can lead to clinically significant problems by missing the opportunity to provide necessary treatment in time. This error type is implemented by detecting if the report describes a positive class for abnormalities and replacing it with a report corresponding to "no finding."

\subsection{Implementation details}
We pre-processed the image data by removing margin space following the approach proposed in a previous work\cite{moon2021multi}. The pre-processing also involved Gaussian blurring, normalization, and resizing the images to 224 $\times$ 224. We used the pre-trained tokenizer of CXR-BERT with a vocabulary size of 30,522\cite{boecking2022making}, and set the maximum word length of the report to 120. For the vision-language pre-training, we employed an AdamW optimizer with an initial and maximum learning rate of 0.00001 and 0.0001, respectively, for five epochs, including two epochs of warm-up period, and the batch size was 10. All experiments were conducted using Python version 3.8 and PyTorch library version 1.10 on two NVIDIA GeForce RTX 3090.


\subsection{Details of evaluation.}
To evaluate the model performances for the zero-shot oversight tasks, we assessed the model in two aspects: zero-shot abnormality detection and zero-shot error detection. As zero-shot abnormality detection can be considered as a zero-shot classification problem, we assessed the zero-shot classification performance of the trained model using a subset of the CheXpert test dataset (500 samples). Following previous works\cite{tiu2022expert}, we calculated the area under the receiver operating curve (AUC) and F1-score for five major abnormalities (atelectasis, cardiomegaly, consoldiation, pulmonary edema, and pleural effusion) and compared the model performance with the current SOTA and other self-supervised and vision-language models.
To calculate the probability of each abnormality class, we performed softmax evaluation for both the simple prompt method ({abnormality} and no {abnormality}) following previous work and the detailed prompt method for each abnormality. The descriptions for each abnormality used in the detailed prompt were five sentences for each class confirmed by board-certified radiologists in the previous paper\cite{zhang2022contrastive}, and new detailed prompts consisting of five sentences each were created by clinicians for the consolidation class for which no corresponding sentences were available. For example, the detailed prompt for "Pleural effusion" includes the following sentences: "A pleural effusion is present.", "Blunting of the costophrenic angles represents pleural effusions.", "Trace pleural fluid is present.", "The pleural space is partially filled with fluid.", and "Layering pleural effusions are present." By performing zero-shot detection using these detailed prompts, we aimed to evaluate how well the model understands detailed expressions beyond simple class names.

We utilized the COVIDx\cite{COVIDx} dataset to evaluate the detection performance of the model on unseen diseases during the learning process of vision-language model. We randomly split 2,998 images from a total of 29,986 images and used them as a test set for evaluation. The images from the COVIDx dataset that were not included in the test set were not utilized during the training process.

We evaluated zero-shot error detection performance on the test set of MIMIC-CXR data that was not used for training, by calculating the AUC and F1-score for detection performance using simulated human errors of five types with a probability of 5\%, as mentioned above.


\subsection{Details of statistical analysis}
For statistical analysis, we used non-parametric bootstrapping. Random samples of the same size as the original dataset were repeatedly sampled with replacement 1,000 times. We estimated the difference in AUC and F1 metrics using the bootstrap samples. Confidence intervals were derived from the relative frequency distribution of the estimates over the re-samples, using the interval between the $100 \times (\alpha/2)$ and $100 \times (1 - \alpha/2)$ percentiles. We set $\alpha$ to 0.05 and considered differences beyond the confidence interval as statistically significant difference.

\subsection{Ethic committee approval.}
The abdominal radiograph data collected for this study were ethically approved by the Institutional Review Boards of Chung-Ang University Hospital, Chungnam University Hospital, and the requirement for informed consent was waived.

 \begin{addendum}
{\color{black} \item[Correspondence] Correspondence and requests for materials should be addressed to Jong Chul Ye.~(email: jong.ye@kaist.ac.kr).}
 \item 
 This research was supported by the KAIST Key Research Institute (Interdisciplinary Research Group) Project, the National Research Foundation of Korea under Grant NRF-2020R1A2B5B03001980, and Chungnam National University Hospital Research Fund, 2022.
{
\item[Author Contributions] S.P. performed all experiments, wrote the extended code, and prepared the manuscript. E.S.L and J.E.L collected data and provided clinical evaluation. J.C.Y. supervised the project in conception and discussion, and prepared the manuscript.}
 \item[Competing Interests] 
The authors declare that they have no competing financial interests.

\item[Data Availability]
Part of the data is collected from open-sourced data repositories that are publicly available. The MIMIC-CXR database is available at \href{https://physionet.org/content/mimic-cxr/2.0.0/}{https://physionet.org/content/mimic-cxr/2.0.0/}. 
Subset of the CheXpert test data and corresponidng labels used for the evaluation of the model in zero-shot abnormality detection can be found at \href{https://github.com/rajpurkarlab/cheXpert-test-set-labels}{https://github.com/rajpurkarlab/cheXpert-test-set-labels}.
COVIDx dataset used for the evauation of the model in unseen disease is available tat \href{https://www.kaggle.com/datasets/andyczhao/covidx-cxr2}{https://www.kaggle.com/datasets/andyczhao/covidx-cxr2}.
Other parts of data used for the experiments for abdominal radiographs are not publicly available due to the patient privacy obligation. Interested users can request access to these data for research purposes, by contacting the corresponding author J.C.Y (jong.ye@kaist.ac.kr). The data can be shared after the IRB approval and de-identification along with the signed agreement on data transfer and usage. Replies to the initial request will be made within 10 working days. The use of data is limited only to the research purpose, and the redistribution is prohibited.
  
\item[Code Availability] 
Th code is available at following GitHub repository. \href{https://github.com/sangjoon-park/Medical_X-VL}{$https://github.com/sangjoon-park/Medical_X-VL$}

\end{addendum}

\section*{References}
\bibliographystyle{naturemag}
\bibliography{ref}

\nolinenumbers

\newpage
\renewcommand{\figurename}{Supplementary Fig.}
\renewcommand{\tablename}{Supplementary Table}

\section*{\huge Supplementary Information}
\par\noindent\rule{\textwidth}{0.5pt}
\vspace{0.75cm}
\\
\setcounter{figure}{0}
\setcounter{table}{0}
\renewcommand{\thefigure}{S\arabic{figure}}

\paragraph{Definition of Momentum Distillation}
The learning paradigm of knowledge distillation involves transferring knowledge from a teacher model to a student model, which was previously explained in the context of self-training. Initially developed for compressing models, knowledge distillation is used to efficiently build a simpler student model by distilling the knowledge from a more complex teacher model. This approach is particularly useful for practical implementation of AI models in devices with limited computational resources. Moreover, knowledge distillation can be employed in a siamese design, where one model learns from the other model's predictions instead of relying on labels. Thus, several works on semi- and self-supervised learning have utilized knowledge distillation in self-training. Recent studies suggest that semi- or self-supervised learning methods based on the knowledge distillation framework can result in a model with performance similar to, or even better than, fully supervised models.
Momentum distillation refers to a type of knowledge distillation where the teacher model, which has the same architecture as the student model, is updated gradually through exponential moving averages from the student model. This approach enables momentum-based updates to be transferred from the teacher to the student model during knowledge distillation.

\vspace{1.5cm}

\begin{table}[htbp]
  \centering
  \caption{Ablation study of the key component of medical X-VL model}
  \begin{adjustbox}{width=0.55\textwidth}
    \begin{tabular}{lp{7.4em}}
    \toprule
    \textbf{} & \multicolumn{1}{l}{\textbf{Average}} \\
    \midrule
        \textbf{AUC} & \multicolumn{1}{c}{} \\
    Proposed (X-VL)  & 0.881 \newline{}(0.843-0.912)  \\
    No MLM  & 0.872 \newline{}(0.834-0.906)  \\
    No momentum distillation  & 0.873 \newline{}(0.835-0.906)  \\
    No sentencewise contrastive learning  & 0.868 \newline{}(0.827-0.904)  \\
    No similarity constraint \newline{} for negative sampling & 0.869 \newline{}(0.829-0.903)  \\
    \bottomrule
                   \multicolumn{2}{l}{\footnotesize  Results are obtained using detailed prompt for five abnormality classes.}
    \end{tabular}%
    \end{adjustbox}
  \label{tab:ablation}%
\end{table}%

\newpage



\begin{figure}[!h]
	\centering
 \centerline{\epsfig{figure=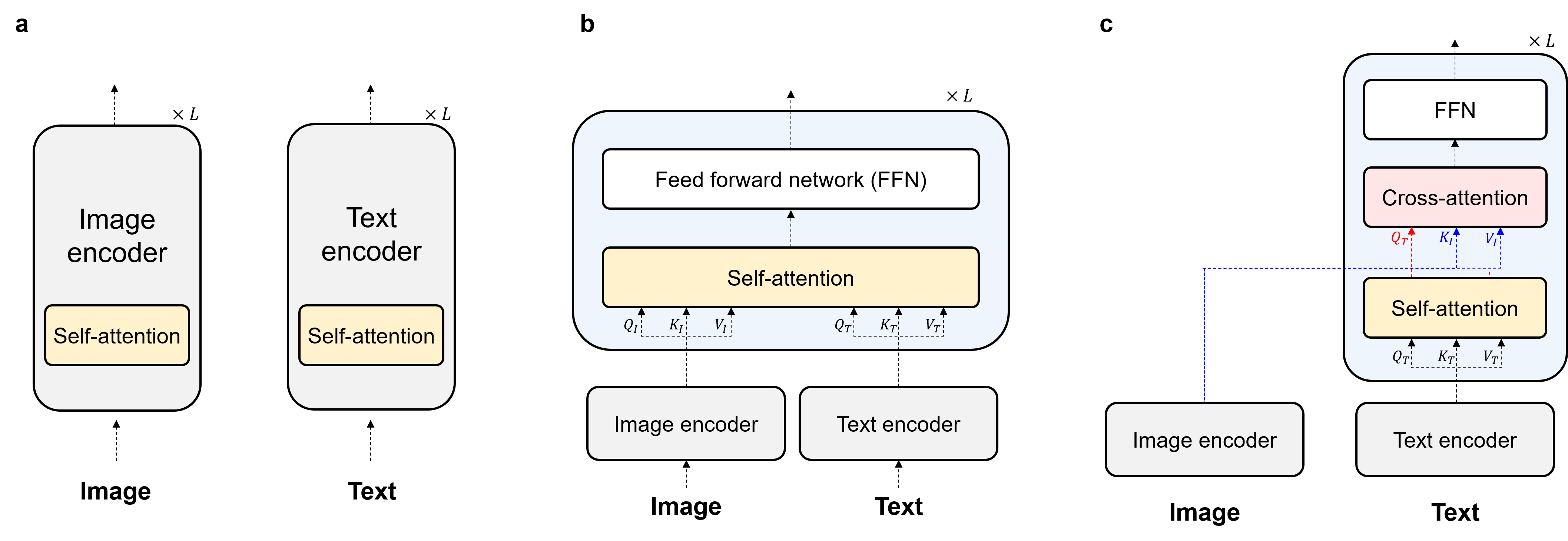, width=1.0\linewidth}}
	\caption{\normalsize Illustration of the different vision-language model architectures. (A) Parallel dual encoder model. Multi-modal fusion encoder model with (B) self-attention and (B) cross-attention.
}
	\label{supple_arch}
\end{figure}

\vspace{2cm}


\begin{figure}[!h]
	\centering
 \centerline{\epsfig{figure=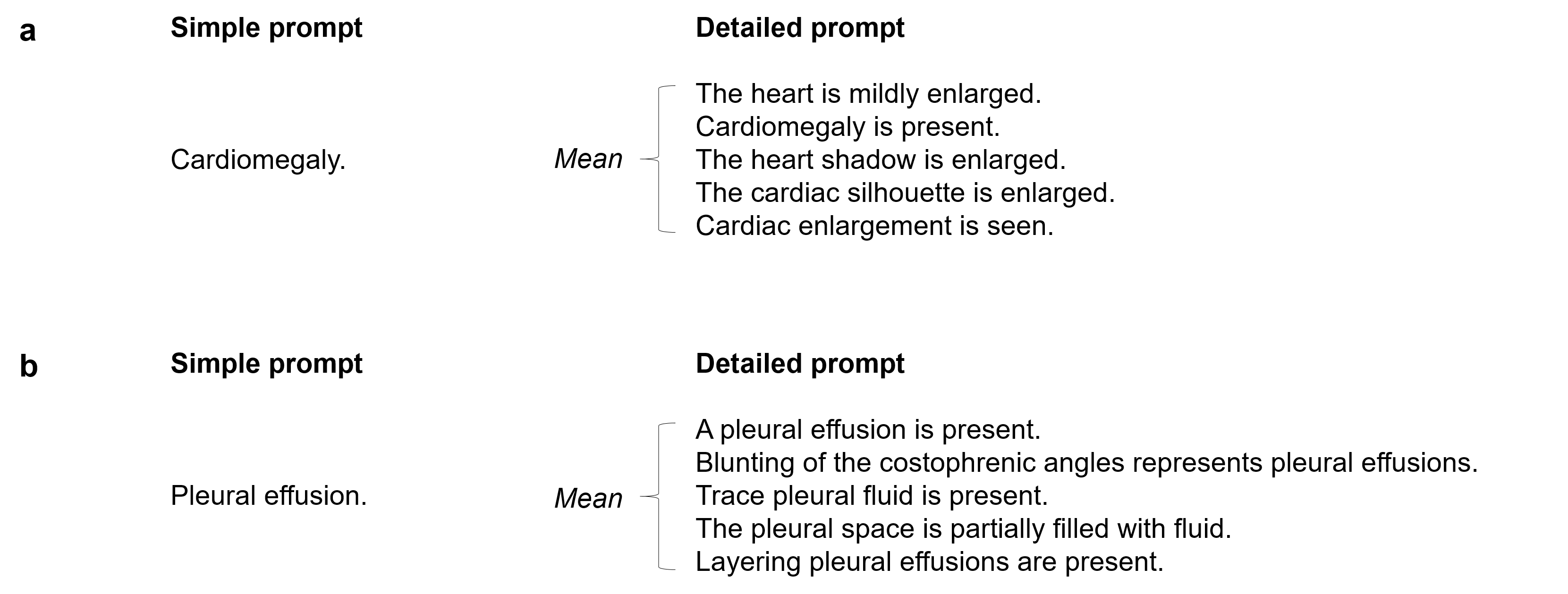, width=1.0\linewidth}}
	\caption{\normalsize Examples of the simple prompt (left) and the detailed prompt (right) for abnormality class (A) cardiomegaly and (B) pleural effusion. For the detailed prompt, evaluation is performed by calculating the mean of the logits of each description.
}
	\label{supple_detailed}
\end{figure}

\begin{figure}[!h]
	\centering
 \centerline{\epsfig{figure=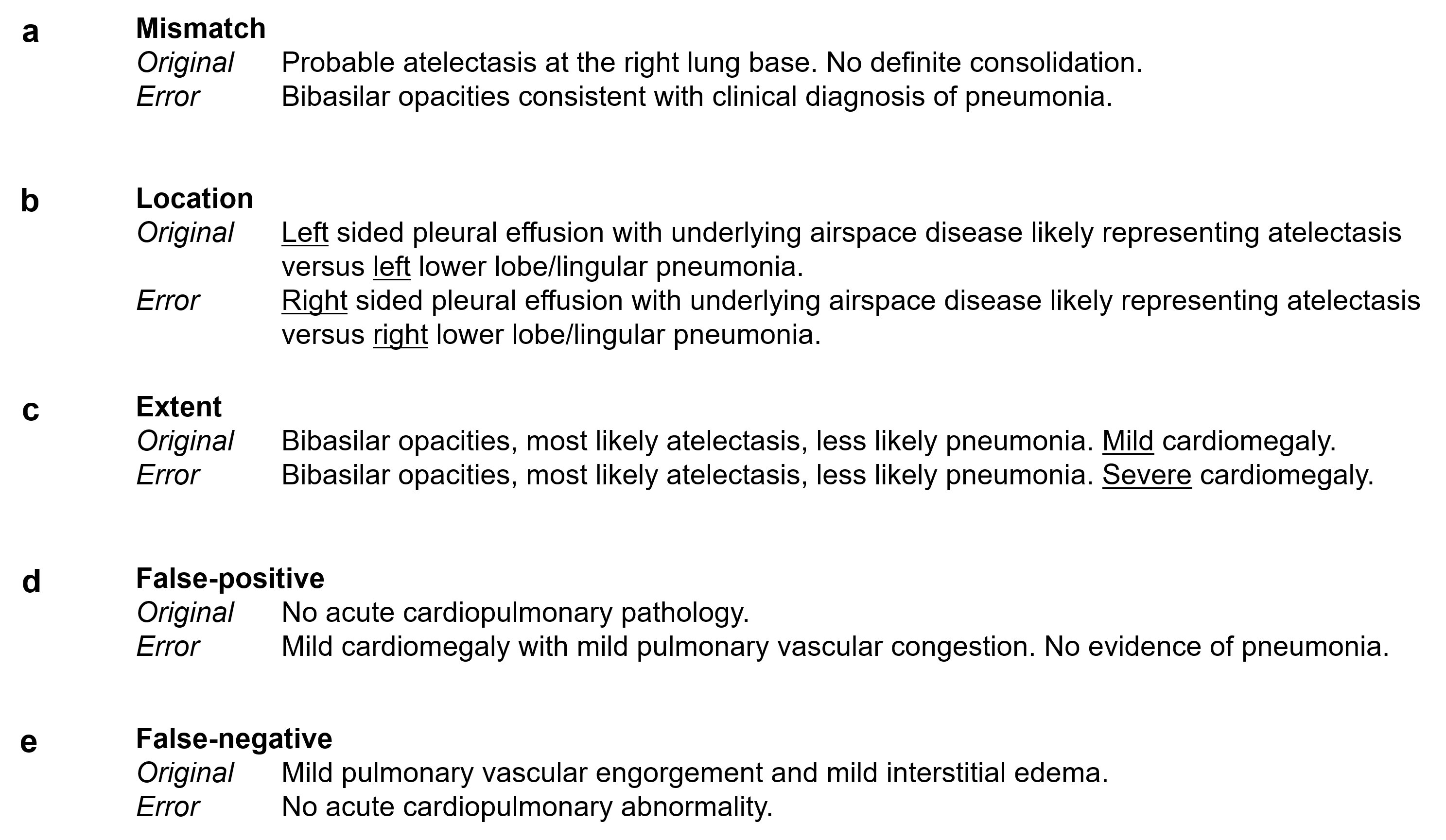, width=1.0\linewidth}}
	\caption{\normalsize Examples of (A) the mismatch, (B) the location, (C) the extent, (D) the false-positive, and (E) the false-negative errors.
}
	\label{supple_report}
\end{figure}

\clearpage
\newpage

\end{document}